\newif\ifpdf
\ifx\pdfoutput\undefined
    \pdffalse
\else
    \pdfoutput=1
    \pdffalse
\fi
\documentclass[preprint,prl,showpacs]{revtex4}
\ifpdf
    \usepackage[pdftex]{graphicx}
    \usepackage[pdftex]{hyperref}
\else
    \usepackage{graphicx}
    \usepackage{hyperref}
\fi
\begin{document}
\preprint{PC-17/ISS 2000 ; October 16, 2000}
\title{Static hole in a critical antiferromagnet: \\
field-theoretic renormalization group}
\author{Subir Sachdev}
\homepage{http://pantheon.yale.edu/~subir}
\affiliation{Department of Physics, Yale University\\
P.O. Box 208120, New Haven, CT 06520-8120, USA}
\begin{abstract}
We consider the quantum field theory of a single, immobile, spin
$S$ hole coupled to a two-dimensional antiferromagnet at a bulk
quantum critical point between phases with and without magnetic
long-range order. We present an alternative derivation of its
two-loop beta function; the results agree completely with earlier
work (M. Vojta {\em et al}, Phys. Rev. B {\bf 61}, 15152 (2000)),
and also determine a new anomalous dimension of the hole creation
operator.

~

~\\ Correspondence: Subir Sachdev, Department of Physics, Yale
University, P.O. Box 208120, New Haven, CT 06520-8120, USA. Fax:
1-203-432-6175. Email: subir.sachdev@yale.edu\\
~\\
{\sc Proceedings of the 13th International Symposium on
Superconductivity, October 14-16, 2000, Tokyo, Japan. To appear
in Physica C.}
\end{abstract}
 \maketitle

Recent papers \cite{science,qimp} have introduced the following
model Hamiltonian for a  single non-magnetic (Zn or Li) impurity
in a two-dimensional $d$-wave superconductor or spin-gap
insulator (see \cite{icmp} for a review and experimental
motivation):
\begin{eqnarray}
{\cal H} &=& {\cal H}_{\phi} - \gamma_0 \hat{S}_{\alpha}
\phi_{\alpha} (x=0) \nonumber \\
{\cal H}_{\phi} &=& \int d^d x \left[ \frac{ \pi_{\alpha}^2 + c^2
(\nabla \phi_{\alpha})^2 + s\phi_{\alpha}^2}{2} + \frac{g_0}{4!}
(\phi_{\alpha}^2)^2 \right]. \label{e1}
\end{eqnarray}
We have written the Hamiltonian in $d$ spatial dimensions, and
$\hat{S}_{\alpha}$ ($\alpha=1,2,3$) are spin $S$ operators of a
magnetic moment that is postulated to be present near the
impurity (the case of physical interest has $S=1/2$); these
operators obey the SU(2) commutation relations
\begin{equation}
[\hat{S}_{\alpha} , \hat{S}_{\beta}] = i
\epsilon_{\alpha\beta\gamma} \hat{S}_{\gamma}
\end{equation}
and $\hat{S}_{\alpha} \hat{S}_{\alpha} = S(S+1)$.
 The field $\phi_{\alpha} (x,t)$
represents the local orientation of the antiferromagnetic order
parameter at spatial position $x$ and time $t$; its canonically
conjugate momentum is $\pi_{\alpha} (x,t)$, and hence
\begin{equation}
[\phi_{\alpha} (x,t), \pi_{\beta} (x',t)] = i \delta_{\alpha
\beta} \delta^d (x-x')
\end{equation}

This theory has a bulk quantum critical point at $s=s_c$ between
a phase with magnetic order ($s<s_c$, $\langle \phi_{\alpha}
\rangle \neq 0$), and a symmetric phase with a spin gap ($s>s_c$,
$\langle \phi_{\alpha} \rangle = 0$). We are interested in the
spin correlations of ${\cal H}$ for $s$ close to $s_c$, and in
the vicinity of the impurity at $x=0$. As discussed in
\cite{science,qimp}, universal aspects of these correlations are
associated with a renormalized continuum theory of ${\cal H}$
defined in an expansion in $\epsilon=3-d$. This renormalization
involves the familiar bulk renormalizations which are insensitive
to the impurity degree of freedom
\begin{equation}
\phi_{\alpha} = \sqrt{Z} \phi_{R \alpha}~~~~;~~~~g_0 =
\frac{\mu^{\epsilon} Z_4}{Z^2 S_{d+1}} g
\end{equation}
and new `boundary' renormalizations associated with the impurity
spin
\begin{equation}
\hat{S}_{\alpha} = \sqrt{Z^{\prime}} \hat{S}_{R
\alpha}~~~~;~~~~\gamma_0 = \frac{\mu^{\epsilon/2}
Z_{\gamma}}{\sqrt{Z Z^{\prime} \widetilde{S}_{d+1}}} \gamma.
\label{e3}
\end{equation}
Here $\mu$ is a renormalization momentum scale (we set the
velocity $c=1$), $S_d = 2/[\Gamma(d/2) (4 \pi)^{d/2}]$, and
$\widetilde{S}_d = \Gamma(d/2-1)/[4 \pi^{d/2}]$. The
renormalization constants $Z$, $Z_4$ were computed long ago
\cite{bgz}; their values in the minimal subtraction scheme to
order $g^2$ are
\begin{equation}
Z = 1 - \frac{5 g^2}{144 \epsilon}~~~;~~~Z_4 = 1 + \frac{11g}{6
\epsilon} + \left(\frac{121}{36 \epsilon^2} - \frac{37}{36
\epsilon} \right)g^2 . \label{e3a}
\end{equation}
The boundary renormalizations were computed to the same order in
\cite{science,qimp}:
\begin{equation}
Z^{\prime} = 1 - \frac{2 \gamma^2}{\epsilon} +
\frac{\gamma^4}{\epsilon}~~~;~~~Z_{\gamma} = 1 +
\frac{\pi^2[S(S+1)-1/3]}{6 \epsilon} \gamma^2 g \label{e5}
\end{equation}

This paper will rederive the above results by a new method which
also yields a renormalization constant for the hole creation
operator. Furthermore, the present approach, unlike that of
\cite{qimp}, has the advantage of being formulated entirely in
terms of perturbation expansion which has a Wick theorem, and can
thus be presented in conventional time-ordered Feynman diagrams.

We will identify the spin $\hat{S}_{\alpha}$ with that of a hole,
with creation operator $\psi^{\dagger}_a$, that has been injected
into the antiferromagnet. So
\begin{equation}
\hat{S}_{\alpha} = \psi^{\dagger}_a L^{\alpha}_{ab} \psi_b
\label{e2}
\end{equation}
where $a,b$ take the $2S+1$ values $-S, \ldots S$, and the
$L^{\alpha}$ are the $(2S+1)\times(2S+1)$ angular momentum
matrices associated with the spin $S$ representation. The hole
operators obey the anticommutation relation
\begin{equation}
\psi^{\dagger}_a \psi_b + \psi_b \psi^{\dagger}_a = \delta_{ab}
\end{equation}
So the remainder of this paper will consider the Hamiltonian
\begin{equation}
{\cal H}_{\psi} = \lambda \psi^{\dagger}_a \psi_a + {\cal
H}_{\phi} - \gamma_0 \psi^{\dagger}_a L^{\alpha}_{ab} \psi_b
\phi_{\alpha} (x=0)
\end{equation}
We will only look at the Hilbert space with a single hole, and
$\lambda$, the energy of this hole is an arbitrary positive
number.

We now consider the renormalization of ${\cal H}_{\psi}$. The
standard procedure suggests the parameterization
\begin{equation}
\psi_a = \sqrt{Z_h} \psi_{Ra}~~~~;~~~~\gamma_0 =
\frac{\mu^{\epsilon/2} \widetilde{Z}_{\gamma}}{Z_h \sqrt{Z
\widetilde{S}_{d+1}}} \gamma. \label{e4}
\end{equation}
It is important to note that despite the relation (\ref{e2}), the
renormalization of the spin $\hat{S}_{\alpha}$ is not the square
of the renormalization of $\psi_a$, $Z^{\prime} \neq Z_h^2$;
bringing the two Fermi operators to the same spacetime point
introduces a composite operator renormalization which invalidates
such a relation. Instead, the relationship between the two
renormalization schemes emerges by comparing the renormalization
of $\gamma_0$ in (\ref{e3}) and (\ref{e4}); consistency of these
relations demands
\begin{equation}
Z_h^2 Z_{\gamma}^2 = \widetilde{Z}_{\gamma}^2 Z^{\prime}
\label{e6}
\end{equation}
We will now compute $Z_h$ and $\widetilde{Z}_{\gamma}$ by
completely standard field theoretic methods, and verify that their
values and (\ref{e5}) satisfy (\ref{e6}).

\begin{figure}
\centerline{\includegraphics[width=4in]{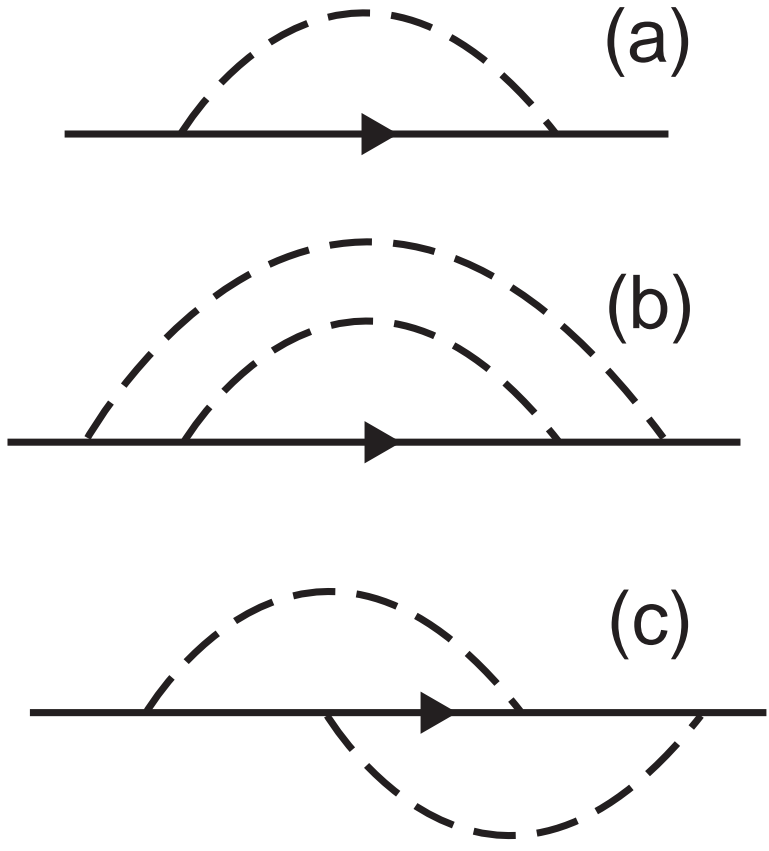}} \vspace{0.5in}
\caption{Diagrams contributing to the $\psi$ fermion self energy.
The full line is the fermion propagator, while the dashed line is
the $\phi_{\alpha}$ propagator.} \label{fig1}
\end{figure}

The Feynman diagrams for the renormalization of two-point $\psi$
Green's function are shown in Fig~\ref{fig1}. As an explicit
example, we display the computation of the simplest one-loop
graph in Fig~\ref{fig1}a:
\begin{eqnarray}
(1a) &=& \gamma_0^2 S(S+1) \int \frac{d^d k}{(2 \pi)^d} \int
\frac{d \omega^{\prime}}{2 \pi} \frac{1}{( \omega^{\prime 2} +
k^2)}
\frac{1}{(- i ( \omega + \omega^{\prime}) + \lambda)} \nonumber \\
&=& \gamma_0^2 S(S+1) \frac{S_d}{2} \int_0^{\infty} \frac{k^{d-2}
d k}{(-i \omega + k + \lambda)} \nonumber \\
&=& A_{\mu} (-i \omega+ \lambda) \gamma^2 S(S+1) \left[ -
\frac{1}{\epsilon} + \aleph/2 + {\cal O}(\epsilon) \right],
\label{r1}
\end{eqnarray}
where $A_{\mu} \equiv \mu^{\epsilon} (- i
\omega+\lambda)^{-\epsilon} \widetilde{Z}_{\gamma}^2 /(Z_h^2 Z)$.
In the last step, the integral was evaluated in dimensional
regularization. The constant $\aleph = -0.8455686701969\ldots$ is
a consequence of phase space factors and will eventually cancel
out of our final results. The remaining diagrams can be evaluated
in a very similar manner: the frequency integrals are performed
first, followed by integrals over the radial momenta. The results
for the two-loop diagrams in Fig~\ref{fig1} are
\begin{eqnarray}
(1b) &=& A_{\mu}^2  (-i \omega+ \lambda) \gamma^4 S^2 (S+1)^2
\left[ \frac{1}{2 \epsilon^2} + \frac{1-\aleph}{2 \epsilon} +
{\cal O}(\epsilon^0) \right]
\nonumber \\
(1c) &=& A_{\mu}^2  (-i \omega+ \lambda) \gamma^4 S (S+1)(S^2 +
S-1) \left[ -\frac{1}{\epsilon^2} + \frac{-1+2\aleph}{2 \epsilon}
+ {\cal O}(\epsilon^0) \right] \label{r2}
\end{eqnarray}

\begin{figure}
\centerline{\includegraphics[width=4in]{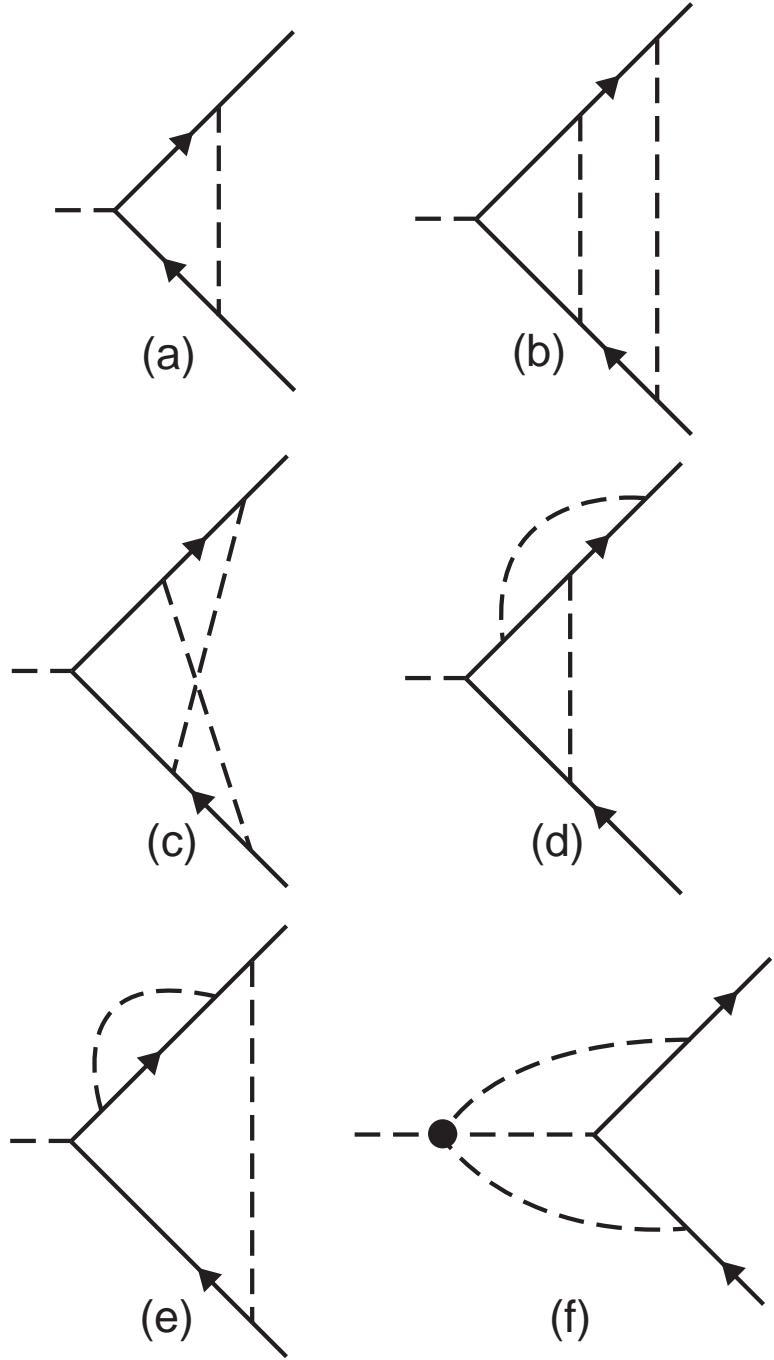}} \vspace{0.5in}
\caption{Diagrams contributing to the renormalization of the
coupling $\gamma$. The full circle is the interaction $g$}
\label{fig2}
\end{figure}

Turning to the renormalization of the vertex $\gamma_0$, the
Feynman diagrams are shown in Fig~\ref{fig2}. Evaluating these as
above we obtain
\begin{eqnarray}
(2a) &=& \gamma_0 A_{\mu} \gamma^2 (S^2 + S -1) \left[
\frac{1}{\epsilon} - 1 - \aleph/2 + {\cal O}(\epsilon)
\right] \nonumber \\
(2b) &=& \gamma_0 A_{\mu}^2 \gamma^4 (S^2 + S -1)^2 \left[
\frac{1}{2\epsilon^2} - \frac{3+\aleph}{2 \epsilon} + {\cal
O}(\epsilon^0) \right]
\nonumber \\
(2c) &=& \gamma_0 A_{\mu}^2 \gamma^4 (S-1)(S+2)(S^2 + S -1) \left[
\frac{1}{2\epsilon} + {\cal O}(\epsilon^0)
\right]\nonumber \\
(2d) &=& \gamma_0 A_{\mu}^2 \gamma^4 (S^2 + S -1)^2 \left[
\frac{1}{\epsilon^2} - \frac{2+\aleph}{2 \epsilon} + {\cal
O}(\epsilon^0) \right]
\nonumber \\
(2e) &=& \gamma_0 A_{\mu}^2 \gamma^4 S(S+1)(S^2 + S -1) \left[
-\frac{1}{\epsilon^2} + \frac{2+\aleph}{2 \epsilon} + {\cal
O}(\epsilon^0) \right]
\nonumber \\
(2f) &=& - \gamma_0 \frac{ A_{\mu}^2 Z_h^2
Z_4}{\widetilde{Z}_{\gamma}^2 Z} \gamma^2 g (S^2 + S -1/3) \left[
\frac{\pi^2}{6\epsilon}  + {\cal O}(\epsilon^0) \right]
 \label{r3}
\end{eqnarray}

The two-loop expression for the boundary renormalization
constants follows immediately from the results
(\ref{r1},\ref{r2},\ref{r3}). Demanding cancellation of poles in
$\epsilon$ in the expressions for the renormalized vertex and
$\psi$ Green's function at external frequency $-i \omega +
\lambda = \mu$ we obtain
\begin{eqnarray}
Z_h &=& 1 - \gamma^2 \frac{S(S+1)}{\epsilon} + \gamma^4 \left[
\frac{(S-1)S(S+1)(S+2)}{2 \epsilon^2} + \frac{S(S+1)}{2 \epsilon}
\right] \nonumber \\
\widetilde{Z}_{\gamma} &=& 1 - \gamma^2 \frac{(S^2 +S -1)
}{\epsilon} + \gamma^4 \left[ \frac{(S^2 + S -3)(S^2+S-1)}{2
\epsilon^2} + \frac{(S^2 + S -1)}{2 \epsilon} \right] \nonumber
\\&~&~~~~~~~~~~~+ g \gamma^2 \frac{\pi^2 ( S^2 +S - 1/3)}{6 \epsilon} \label{r4}
\end{eqnarray}
It can be checked that (\ref{r4}) and (\ref{e5}) satisfy
(\ref{e6}).

The validity of (\ref{e6}) implies that the beta function for the
coupling $\gamma$ is the same as that in \cite{qimp}. Using either
(\ref{e3},\ref{e3a},\ref{e5}) or (\ref{e4},\ref{e3a},\ref{r4}) we
obtain
\begin{equation}
\beta(\gamma) = - \frac{\epsilon \gamma}{2}+\gamma^3 -\gamma^5 +
\frac{5 g^2 \gamma}{144} + \frac{g \gamma^3 \pi^2}{3} (S^2 +
S-1/3). \label{beta1}
\end{equation}
The anomalous dimension of the $\psi_a$ field at the quantum
critical point also follows from (\ref{r4})
\begin{equation}
\eta_h = \beta (\gamma) \frac{d \ln Z_h}{d \gamma} = S(S+1) (
\gamma^2 - \gamma^4 ), \label{r5}
\end{equation}
while, as in \cite{qimp}, the anomalous dimension of the spin
field, $\hat{S}_{\alpha}$, follows from (\ref{e3},\ref{e5}):
\begin{equation}
\eta^{\prime}=2(\gamma^2 - \gamma^4). \label{r5a}
\end{equation}
For completeness we also note the
beta function for the coupling $g$ which follows from (\ref{e3a})
\begin{equation}
\beta(g) = -\epsilon g + \frac{11g^2}{6} - \frac{23g^3}{12}.
\label{beta2}
\end{equation}

The stable fixed point of the beta functions
(\ref{beta1},\ref{beta2}) has $g \neq 0$ and $\gamma \neq 0$
\cite{qimp}. Evaluating (\ref{r5}) at the fixed point of the beta
functions \cite{qimp}, we obtain
\begin{equation}
\eta_h = S(S+1) \left[ \frac{\epsilon}{2} - \left( \frac{5}{484}
+ \frac{\pi^2(S^2 +S -1/3)}{11} \right) \epsilon^2 + {\cal
O}(\epsilon^3) \right] \label{r6}
\end{equation}
($\eta^{\prime} = 2\eta_h/[S(S+1)]$ at this order). This
anomalous dimension implies that the Green's function $G =
\langle \psi_a \psi_a^{\dagger} \rangle$ obeys
\begin{equation}
G (\omega) \sim (\lambda-\omega)^{-1+\eta_h}.
\end{equation}
 The equations
(\ref{r4},\ref{r5},\ref{r6}) are the main new results of this
paper. Unfortunately, the order $\epsilon^2$ corrections in
(\ref{r6}) are rather large: this suggests that truncating the
asymptotic series for $\eta_h$ at order $\epsilon$ probably gives
the most reasonable estimate for its numerical value.

There is also an unstable fixed point at which the bulk
interactions vanish ($g=0$). As shown in \cite{qimp},
$\eta'=\epsilon$ exactly at this fixed point, and here we find
that $\eta_h = S(S+1) \epsilon/2 + {\cal O}(\epsilon^3)$. There
appears to be no general reason for the higher order terms in
$\eta_h$ to vanish. The $g=0$ fixed point can also be studied in
a large $N$ theory \cite{qimp}, and the $N=\infty$ results are
$\eta'=1$ and $\eta_h=1/2$.

The physical motivations and implications of the above results
are discussed in a separate paper \cite{stv}: there we argue that
the anomalous dimension $\eta_h$ characterizes photoemission
spectra of {\em mobile} holes in two-dimensional antiferromagnets
and superconductors in the vicinity of points in the Brillouin
zone where their dispersion spectra are quadratic ({\em i.e.}
near energy minima, maxima, and van Hove singularities). The
intensively studied $(\pi, 0)$, $(0,\pi)$ points (the anti-nodal
points) in the high temperature superconductors are prominent
examples.

\begin{acknowledgements}
This research was supported by US NSF Grant No DMR 96--23181.

\end{acknowledgements}

\end{document}